\documentclass[11pt]{article}
\usepackage{setspace,braket}
\usepackage{amsmath, amssymb, amsfonts, bm}
\usepackage{jheppub}
\def\be{\begin{equation}}
\def\ee{\end{equation}}

\def\H{{\mathcal H}}

\def\be{\begin{equation}}
\def\ee{\end{equation}}

\def\bg{\bar{g}}
\def\beq{\begin{eqnarray}}\def\eeq{\end{eqnarray}}
\def\ba#1\ea{\begin{align}#1\end{align}}
\def\bg#1\eg{\begin{gather}#1\end{gather}}
\def\bm#1\em{\begin{multline}#1\end{multline}}
\def\bmd#1\emd{\begin{multlined}#1\end{multlined}}

\def\nn{\nonumber}

\def\({\left(}
\def\){\right)}
\def\[{\left[}
\def\]{\right]}

\newcommand{\bea}{\begin{eqnarray}}
\newcommand{\eea}{\end{eqnarray}}

\begin{document}

\preprint{WITS-CTP-141}

\title{Attractive holographic $c$-functions }
\author[a]{Arpan Bhattacharyya,}
\author[b]{S.\ Shajidul Haque,}
\author[b]{Vishnu Jejjala,}
\author[b]{Suresh Nampuri,}
\author[b]{and \'{A}lvaro V\'{e}liz-Osorio}
\affiliation[a]{Centre for High Energy Physics, Indian Institute of Science, C.V.\ Raman Avenue, Bangalore 560012, India}
\affiliation[b]{Centre for Theoretical Physics, NITheP, and School of Physics,
University of the Witwatersrand,\\ Johannesburg, WITS 2050, South Africa}
\emailAdd{arpan@cts.iisc.ernet.in}
\emailAdd{Shajid.Haque@wits.ac.za}
\emailAdd{vishnu@neo.phys.wits.ac.za}
\emailAdd{Suresh.Nampuri@wits.ac.za}
\emailAdd{Alvaro.VelizOsorio@wits.ac.za}
\begin{abstract}{
Using the attractor mechanism for extremal solutions in ${\cal N}=2$ gauged supergravity, we construct a $c$-function that interpolates between the central charges of theories at ultraviolet and infrared conformal fixed points corresponding to anti-de Sitter geometries.
The $c$-function we obtain is couched purely in terms of bulk quantities and connects two different dimensional CFTs at the stable conformal fixed points under the RG flow.}
\end{abstract}
\maketitle

\section{Introduction}
When we integrate out ultraviolet degrees of freedom to obtain a low-energy effective theory, we inevitably lose information about short distance physics.
The degrees of freedom become coarse-grained in the infrared.
We would like to understand how this works in detail in a gravitational setting.
Extremal black holes in anti-de Sitter (AdS) space supply a useful laboratory for investigating what happens to the gravitational degrees of freedom as we perform the integration.
Near to the horizon of an extremal black hole in AdS$_5$, for example, the geometry is AdS$_2 \times S^3$.
Generically, the dimensions of the AdS spaces at the asymptopia and the near-horizon region are different.
The duality between string theory on AdS geometries and the conformal field theory (CFT) resident at the boundary of the spacetime~\cite{m,gkp,w} allows us to examine the gravitational degrees of freedom carefully because the entropy of the black hole is accounted for both by the enumeration of states in the CFT corresponding to the asymptotic region and in the CFT corresponding to the near-horizon region.
We should as well see that the overall number of degrees of freedom decreases as we migrate to the interior of the spacetime.
The $c$-theorem enables us to quantify the difference between the central charges of the two fixed point theories at either end of the flow.
In this paper, we wish to make this statement precise in non-trivial supergravity environments with matter fields for which the dimension of the AdS factor changes as we traverse the radial direction in the bulk toward an extremal horizon.
We will do this by providing an algorithm for constructing a $c$-function.

In investigating critical models in two dimensions, Zamolodchikov proved a remarkable result:
when conformal fixed points are connected by a renormalization group (RG) flow, there exists a positive real function $c(g_i, \Lambda)$ of the coupling constants and energy scale whose value does not increase along the trajectory.
At the fixed points of the flow, where the beta functions vanish, this $c$-function is stationary and assumes values equal to the central charges of the corresponding conformal field theories (CFT)~\cite{zam}.
As $c_\mathrm{UV} \ge c_\mathrm{IR}$, the RG evolution of the $c$-function is a gradient flow in which the number of degrees of freedom decreases.
In crafting a four-dimensional analogue of the $c$-theorem from two dimensions, we compute the vacuum expectation value of the trace of the stress-energy tensor to be
\be
16\pi^2 \langle T^\mu_\mu \rangle = -a\, E_4 + c\, W_{\mu\nu\rho\sigma} W^{\mu\nu\rho\sigma} ~, \label{eulerweyl}
\ee
where $E_4$, the Euler density, is quadratic in the Riemann tensor and $W_{\mu\nu\rho\sigma}$ is the Weyl tensor.
While in all known examples, $a_\mathrm{UV} \ge a_\mathrm{IR}$, the same is not true of the coefficient $c$ in~\eqref{eulerweyl}.
Thus, it is the coefficient of the Euler density that offers a candidate for the $c$-function in higher dimensions~\cite{car,osb1,osb2}, and indeed, the $a$-theorem is now firmly established in four dimensions~\cite{ks,zk}.

The gauge/gravity duality allows us to map the $c$-theorem to a statement about gravity~\cite{gppz,fgpw, 0201270, sinha, sinha1, lsz, paulos}.
In general, the $c$-functions obtained in this manner parameterize an RG flow that starts from the vacuum state in the ultraviolet and terminates in the vacuum state in the infrared.
For spherically symmetric, static configurations in four-dimensional two derivative gravity, a monotonically decreasing function was proposed in~\cite{gjmt} that applies to non-vacuum settings.
This is simply the area of radial slices in the bulk.
At the horizon of the black hole, \textit{i.e.}, in the infrared, the entropy is an $SO(3)$ invariant function that counts the degrees of freedom.
This proposal is, however, unsatisfactory because it diverges on the boundary.

In this paper, we establish that for AdS spaces with supersymmetric single centered black holes, the $c$-theorem arises as a consequence of the attractor mechanism~\cite{attr1,attr1a, attr1b,attr1c,attr2,attr2a,attr2b,attr2c,attr2d}.
At large radius, the geometry is asymptotically AdS$_{d+1}$.
Near the horizon of the black hole (black brane), the metric has an AdS$_3$ or AdS$_2$ factor.
The $c$-function we construct therefore interpolates between AdS spacetimes of different dimensions.
At the endpoints of the flow, the $c$-function is extremized and assumes the values of the central charges corresponding to the AdS geometry in the ultraviolet or the infrared.
Since we work in ${\cal N}=2$ supergravity, the stress-energy tensor in the bulk is highly non-trivial.
Typically scalar fields and fluxes are present.
Because of the power of the attractor mechanism, the prescription for finding the $c$-function that we detail in this paper applies to this setting.
Supersymmetry, as well, is not required.
It is sufficient that the black background under consideration be extremal.
We work in gauged supergravity and neglect higher derivatives and string loop effects, but our analysis is otherwise completely general and provides a recipe for calculating a $c$-function that interpolates between non-vacuum states.

The outline of the paper is as follows.
In Section~\ref{sec1}, we detail the attractor flows that we investigate in this work.
In Section~\ref{sec2}, we develop the form of the $c$-function in supergravity.
The null energy condition ensures that the function is monotonic.
In Section~\ref{thecon}, we write the $c$-functions explicitly for black backgrounds with AdS$_{d+1}$ asymptopia and an AdS$_3$ or an AdS$_2$ factor in the near-horizon region.
In Section~\ref{constants}, we evaluate the example of a black brane in AdS$_5$ with a near-horizon AdS$_3$ factor and explain how to fix the constants in the $c$-function.
In Section~\ref{disc}, we discuss our results and provide a prospectus for future work.

\section{Setting up the evolution}\label{sec1}
In holographic RG, the radial direction in AdS is a proxy for the scale of the Wilsonian flow in the dual field theory~\cite{ktdvv,ktdvva,ktdvvb,ktdvvc}.
In the bulk, we then have a function $c(r) = F(\phi^i(r))$, where $\phi^i(r)$ are scalar fields in the spacetime that are dual to the operators sourced by the coupling constants on the boundary.
In the dual field theory, the $c$-function curve is given by $c(g_i,\Lambda)= F(g_i(\Lambda))$, where $\Lambda$ is the energy scale.
The evolution of the $c$-function in the field theory follows a Callan--Symanzik equation where the choice of how we parameterize energy corresponds to the choice of the bulk radial coordinate.
In describing the $c$-function, we are perfectly free to think of it as a function of $r^2$, for instance.
There is an inbuilt redundancy in the characterization of the $c$-function corresponding to this freedom.
We need to demand only that at the endpoints of the RG flow, the function takes values corresponding to the central charges at the fixed points, where the theory is exactly conformal.
These central charges are related to the radius of the AdS geometry dual to the fixed point.
In between the endpoints, due to the choice of the radial coordinate, the flow can be different, but each of the $c$-functions obtained in this manner is monotonic.

We specialize to ${\cal N}=2$ $U(1)$ gauged supergravity and use the attractor mechanism for black solutions in the bulk~\cite{attr2,attr2a,attr2b,attr2c,attr2d}.
The Hilbert space of the dual field theory is graded by temperature, charge, angular momentum, and flux quantum numbers that also distinguish the black backgrounds in the bulk.
As we consider single centered solutions here, we restrict the field theory to the subspace of the Hilbert space with the appropriate charges and consider the Wilsonian evolution of only this subspace.
In particular, we will not consider solutions with multipole moments or hair.
The states we enumerate in this subspace contribute to the universal large charge limit of the entropy.
In the bulk, these states are recognized as single centered black hole (black brane) microstates, whose number reproduces the Bekenstein--Hawking entropy.
For extremal black holes, the entropy function $S = \frac14 A$ (in units where the Newton constant $G=1$ and $A$ is the area of the horizon) is a fixed quantity.

The near-horizon geometry of the extremal background is of the form AdS$_2\times X$ or AdS$_3 \times X$, where $X$ can be spherical, hyperbolic, or planar depending on the structure of the black hole (black brane) horizon.
The horizon acts as an attractive fixed point for the scalar flows in this background.
An attractive on-shell scalar flow from the asymptotic AdS boundary to the near-horizon AdS factor is holographically dual to the Wilsonian flow of the effective field theory action that encodes the dynamics of the field theory operators dual to the scalar fields in the specified black background.
The flow interpolates between the ultraviolet fixed point CFT corresponding to the asymptotic AdS geometry and the infrared fixed point CFT corresponding to the near-horizon AdS geometry.
The near-horizon AdS geometry is dual either to a two-dimensional CFT or to the discrete light-cone quantization (DLCQ) limit of one~\cite{bdss}.
The subset of states that corresponds to the single centered black solution are encoded in a Virasoro algebra representation of a two-dimensional CFT in the infrared.
The $c$-function that acts as an affine parameter for this Wilsonian flow is dual to an affine parametrization of the attractive scalar flows in the bulk.
AdS$_3\times X$ is stable under small perturbations.
Infrared fixed points of the form of AdS$_2 \times X$ have been studied extensively especially in the context of stability under perturbations and the interested reader is referred to~\cite{Jain,Jaina,Jaina1,Jainb,Jainc} for further details.

\section{The nature of the solution} \label{sec2}
In order to focus our discussion, we recall that in four-dimensional ${\cal N}=2$ gauged supergravity in the presence of fluxes that arise from a low-energy string compactification, the gravity multiplet couples to massless gauge fields.
The coupling strengths of these interactions are determined by the scalars that we have mentioned.
While supersymmetry will not ultimately turn out to be essential to the analysis, we concentrate initially on supersymmetric flows in asymptotically AdS, static, spherically symmetric, extremal backgrounds.
Deducing a suitable $c$-function in the bulk is equivalent to finding a real function of the scalar fields which is invariant under symplectic transformations of the scalar moduli space.
We would like to encode this proposal in terms of purely geometrical quantities and then to generalize the result to non-supersymmetric attractive flows in any dimension.
The central charges are read off at both boundaries by computing the renormalized part of the stress-energy tensor.
These terms depend purely on the metric and are independent of the matter content of the theory.
This motivates the idea that something purely geometrical arrests all the information about the central charges at the fixed points of the flow.
The $c$-function must have extremization conditions at these endpoints and monotonicity properties that are independent of the matter content of the field theory and must depend on the Einstein tensor $G_{\mu\nu}$.

To ensure monotonicity of the $c$-function, we examine its derivative.
A geometric function $c(r)$ must have a derivative $c'(r)$ that is proportional to a scalar quantity in gravity constructed purely out of the stress-energy tensor with a definite signature.
(All primes denote differentiation with respect to the radial coordinate $r$.)
A natural choice follows from the null energy condition, which states that for any null vector $k$, $T_{\mu\nu}k^\mu k^\nu \geq 0$ for all physical backgrounds.
The inequality is saturated only in vacuum backgrounds.
This is a physical input for the construction.

In order to deduce a suitable $c$-function, we first observe that the attractive scalar flow can be encoded in terms of a first order equation,
\begin{equation}
{\phi^i}' = f^i(g_{\mu\nu}(r), \phi^i(r)) ~, \label{phip}
\end{equation}
where $\phi^i$s are the scalar fields in the bulk and $g_{\mu\nu}$ represents the bulk metric.
In the static and spherically symmetric backgrounds that we consider, $\phi^i$ is a function only of the radial coordinate $r$.
From~\eqref{phip}, it follows that in the scalar moduli space, a vector valued function $f(r)$ can be defined as a conservative function that vanishes at both endpoints of the flow and can be written as a gradient $f^i(r) = G^{ij}\partial_j \Upsilon$, where $G_{ij}$ is the metric on the scalar space.
Hence, the attractor equation can be rewritten as
\begin{equation}
{\phi^i}' (r)= G^{ij}\partial_j \Upsilon~.\label{phip1}
\end{equation}
Furthermore, in such a first order dynamical system, there is a first order equation which relates the metric coefficients and their derivatives to the function $\Upsilon$ that drives the scalar flow:
\begin{equation}
\Upsilon = p(g_{\mu\nu}(r),g'_{\mu\nu}(r)) ~. \label{at}
\end{equation}

In order for ${\phi^i}'(r)$ to vanish, the function $\Upsilon$ has to be extremized only at the endpoints of the flow.
Hence, a function that serves as an effective affine parameter for the attractive flow in moduli space could simply be proportional to $\Upsilon$ up to a constant:
\begin{equation} \label{at1}
c(\phi^i)= \lambda + \kappa \Upsilon ~.
\end{equation}
The function $\Upsilon$ is proportional to the generalized superpotential in ${\cal N}=2$ theories and in particular becomes proportional to the central charges for supersymmetric attractive flows.
This choice is motivated by the need to define a function which is specified by a minimum number of dynamical conditions and parameters of the system and applies our understanding that along supersymmetric attractive flows the central charge gets minimized.
In order to define the bulk equivalent of the $c$-function for the RG flow in the boundary, we need to write down $c$ as a function of the radial coordinate $r$ dual to the boundary energy scale in terms of bulk geometrical quantities.
The simplest procedure to do this would be to use the attractor equations to encode $\Upsilon$ in terms of the geometry.
From~\eqref{at}, we see that the function $p(r)$ performs this encoding operation and by construction is monotonic along the attractor flow.

We write down a $c$-function that is monotonic as
\beq
c(r) = \lambda + \kappa\,[ \chi(p (r))+\mathcal{G}(r) ] ~, \label{blah}
\eeq
where $\mathcal{G}(r)$ is a differentiable monotonic function and $\chi(p(r))$ is a differentiable function in $p(r)$ which is extremized in scalar moduli space at the endpoints.
We label the term in brackets in~\eqref{blah} as $\H(r)$ for convenience.
Imposing the absence of extremization along the flow, except at the two endpoints, leads us to the constraint that the derivative $\H'(r)$ is of one sign only:
\bea
\H'(r) \geq 0 &\qquad \mathrm{or} \qquad& \H'(r) \leq 0 ~,
\eea
with the inequality saturated only at the endpoints $r=r_h$ and $r=\infty$.

The only scalar quantity that depends on the matter content of the theory and has a definite signature determined purely from the geometry for a physical on-shell background with a null condition achieved on vacuum solutions is a contraction of the stress tensor $T_{\mu\nu}$ with any null vector $k^\mu$:
\be \label{imp0}
8\pi T_{\mu \nu} k^{\mu} k^{\nu} = G_{\mu\nu} k^\mu k^\nu \ge 0 ~.
\ee
Now, we can always write this contraction as
\be \label{imp}
8\pi T_{\mu \nu} k^{\mu} k^{\nu} = \mathcal{F} (g_{\mu\nu})\, \mathcal{B}(g_{\mu\nu}, g'_{\mu\nu}) ~,
\ee
where $\mathcal{F}$ is a positive regular function of the metric and $\mathcal{B}$ is a positive regular function of the metric and its derivative.
For a given $\chi(p(r))$, we calculate $\mathcal{G}(r)$ by defining
\be
\H'(r) = \frac{8\pi T_{\mu \nu} k^{\mu} k^{\nu}}{\mathcal{F}(g_{\mu\nu})} ~.
\ee

It is then manifestly true that $\H'(r)\geq 0$ as a consequence of the null energy condition.
We therefore determine that $\mathcal{H}(r)$, and consequently $c(r)$, is a monotonically decreasing function from the ultraviolet to the infrared.
The null energy condition fixes the behavior of $\mathcal{H}(r)$ completely.
The constants $\lambda$ and $\kappa$ in~\eqref{blah} are determined by matching to the central charges at the two endpoints of the flow.
This is our prescription for deriving the $c$-function.

For a given null vector and $p(r)$, we can in fact choose various $\mathcal {F}(r)$ and $\mathcal{B}(r)$ functions with the same product and hence various corresponding functions $\chi$.
The freedom in choosing the $\mathcal{F}$ function is related to the arbitrariness in picking a $c$-function for RG flows on the boundary depending on the renormalization scheme used.
Furthermore, in going from the central charge to a geometrical representation, there is an extra degree of freedom available in terms of the $SO(d-1,1)$ group of coordinate diffeomorphisms which can operate on the non-radial part of the metric.
We fix the diffeomorphism gauge by choosing the metric to be static and spherically symmetric.
Such a metric can support black backgrounds with an AdS$_2$ factor or an AdS$_3$ factor in the infrared as the endpoint of the attractive flow.

\section{The construction} \label{thecon}
To be explicit, we start with a $(d+1)$-dimensional static spherically symmetric metric of the form
\beq \label{dmetric}
ds^2= -a(r)^2 dt^2+a(r)^{-2} dr^2+ b(r)^2 \sum_{i=1}^{d-2} dx_i^2 + w(r)^2 dz^2~. \label{metric}
\eeq
The coordinate $r$ is the radial direction corresponding to the RG flow, and $t,x_{i}$, and $z$ are boundary coordinates.
The harmonic function $a(r)$ and the warp factors $b(r)$ and $w(r)$ that appear in~\eqref{metric} describe a black hole spacetime in the infrared and give an asymptotically AdS$_{d+1}$ geometry for large $r$.
The horizon of the extremal black hole localizes at $r=r_h$, where $a(r_h)=0$.
The singularity exists at $b(r)=0$.
At the horizon $r=r_h$, the area is written in terms of $b(r_h)\, \sqrt{w(r_h)}$~\cite{bdchn}.
Crucially, the metric~\eqref{metric} allows for a near-horizon geometry with an AdS$_2$ factor when $w(r)= b(r)$ and for an AdS$_3$ factor when $w(r)=a(r)$.

Consider the null vector
\be
k^{\mu} = (a(r)^{-1},a(r),0,\ldots,0) \label{nvec} ~.
\ee
From the Einstein equation for the $(d+1)$-dimensional metric given in~\eqref{dmetric}, the null energy contraction reads
\beq
8\pi T_{\mu\nu}\,k^\mu k^\nu =- a(r)^2\left(\,\frac{w''(r)}{w(r)}+ (d-2)\,\frac{b''(r)}{b(r)}\right)\,\geq 0 ~. \label{Tkk2}
\eeq

\subsection{AdS$_3$ near-horizon geometry} \label{ads3}
We set $w(r)=a(r)$.
The attractor equations tell us that~\cite{bdchn}
\be
p(r) = \frac{a'(r)}{a(r)}+ (d-2)\,\frac{b'(r)}{b(r)} ~.
\ee
Now, keeping in mind~\eqref{imp} and the arguments of Section~\ref{sec2}, we choose $\mathcal{F}(r)= a(r)^2$ and let $\chi(p(r))=-p(r)$.
Then, in order to simultaneously satisfy
\bea
\mathcal{H}(r) = -p(r) + \mathcal{G}(r) & \qquad \mathrm{and} \qquad &
\mathcal{H}'(r) = \frac{ 8\pi T_{\mu \nu}k^{\mu}k^{\nu}}{\mathcal{F}(r)} ~,
\eea
we deduce
\be
\mathcal{G}(r) = - \int dr\, \left\{ \left(\frac{a'(r)}{a(r)}\right)^2+ (d-2) \left(\frac{b'(r)}{b(r)}\right)^2\right\}~.
\ee
Putting the pieces together, the $c$-function can be formally written as
\be \label{c3}
c_{\mathrm{AdS}_3}= \lambda + \kappa\left[ \left(\frac{a'(r)}{a(r)}+(d-2)\frac{b'(r)}{b(r)}
\right) + \int dr\, \left\{\left(\frac{a'(r)}{a(r)}\right)^2 + (d-2)\left (\frac{b'(r)}{b(r)}\right)^2\right \} \right] ~,
\ee
where we have absorbed signs into $\kappa$.
We verify that $c(r)$ as defined in~\eqref{c3} is monotonically non-increasing from the ultraviolet to the infrared as a consequence of~\eqref{Tkk2}.

\subsection{AdS$_2$ near-horizon geometry}
Similarly, using $w(r)=b(r)$ in order to ensure the existence of an AdS$_2$ factor in the near-horizon geometry, we employ the attractor equation~\cite{bchno}
\be
p(r) = (d-1)\,\frac{b'(r)}{b(r)} ~.
\ee
Following the same arguments as in Section~\ref{ads3}, we choose
\bea
\mathcal{F}(r)= (d-1)\, \frac{a(r)^2}{b(r)} ~, & \qquad \qquad &
\chi(p(r))= -\frac{b(r)\, p(r)}{d-1} = -b'(r) ~.
\eea
The relations
\bea
\mathcal{H}(r) = -b'(r) + \mathcal{G}(r) & \qquad \mathrm{and} \qquad &
\mathcal{H}'(r) = \frac{ 8\pi T_{\mu \nu}k^{\mu}k^{\nu}}{\mathcal{F}(r)}
\eea
are satisfied for
\be
\mathcal{G}(r) =0 ~.
\ee
This yields the $c$-function
\beq \label{c2}
c_{\mathrm{AdS}_2}= \lambda + \kappa\, b'(r) ~.
\eeq

The extremization of the $c$-function follows from the fact that $b''(r) = 0$ in all AdS spaces as these geometries saturate the null energy condition.
One can also obtain the $p(r)$ by looking at illustrative examples of interpolating solutions between AdS$_4$ and AdS$_2$ as in~\cite{dg,bchno,avo} as a consistency check.

To compare the expressions~\eqref{at1} and~\eqref{c2}, let us define $\phi^i=  X^i /X^0$,  where $X^I$ are the scalars in the vector multiplets of the ${\cal N}=2$ supergravity theory in four dimensions.
(Note that $i=1,\ldots,n_V$ and $I=0,\ldots,n_V$, with $n_V+1$ the number of vector multiplets.)
Following~\cite{dg,bchno}, we may calculate
\be
b'(r) = (a\, b)^{-1} {\cal Z} (\phi) ~,
\ee
where
\be
{\cal Z} (\phi) =\,|Z(\phi)-ib^2\, W(\phi)|
\ee
is the generalized superpotential, a combination of the central charge $Z$ and the superpotential $W$.
With this structure, we can explicitly verify that $b'(r_h)$ is regular.
The function $\Upsilon$ then is
\be
\Upsilon = b'(r) = \frac{1}{a\, b} \, |Z(\phi)-ib^2\, W(\phi)| ~.
\ee

\subsection{Comments on the $c$-function}
The form of the $c$-function in~(\ref{c3}) and~\eqref{c2} is robust.
This function interpolates between Lorentzian fixed points in the geometry, is written purely in terms of geometrical quantities, \textit{viz.}, the harmonic function and the warp factor in the metric, and satisfies the monotonicity and extremization properties in Zamolodchikov's theorem.
The unknown parameters are determined by the boundary conditions.

The construction in this paper therefore applies to extremal black solutions that interpolate between AdS$_{d+1}$ asymptopia to AdS$_2$ or AdS$_3$ near-horizon regimes.
Examples include (a) extremal black branes in five dimensions, (b) extremal black strings in five dimensions, and (c) BPS black branes and black holes in four dimensions~\cite{hr}:
\bea
\mathrm{a.} &\qquad& \mathrm{AdS}_5 \stackrel{\mathrm{RG}}{\longrightarrow} \mathrm{AdS}_2\times \mathbb{R}^3 ~, \nn \\
\mathrm{b.} &\qquad& \mathrm{AdS}_5 \stackrel{\mathrm{RG}}{\longrightarrow} \mathrm{AdS}_3\times \Sigma_k^2 ~, \label{cases} \\
\mathrm{c.} &\qquad& \mathrm{AdS}_4 \stackrel{\mathrm{RG}}{\longrightarrow} \mathrm{AdS}_2\times \Sigma_k^2 ~. \nn
\eea
Here, $\Sigma_k^2$ refers to two-dimensional flat space ($k=0$), the two-sphere ($k=1$), and hyperbolic space ($k=-1$).
Black holes with two R-charges and three R-charges in AdS$_5$ have, respectively, AdS$_3$ and AdS$_2$ factors in decoupling limits near the horizon~\cite{bdjs}.
Extremal vanishing horizon black holes generically have an AdS$_3$ factor in their near-horizon geometry~\cite{shahin1,sh2,sh3,sh4}.

Case (c) above requires some elaboration.
For AdS$_{4}$, the corresponding CFT is three-dimensional.
For odd dimensional CFTs, the vanishing trace anomaly term implies a vanishing central charge.
In these theories, the free energy of the CFTs conformally mapped to a sphere is proposed to be the monotonically decreasing function that is stationary at the ultraviolet and infrared fixed points.
More specifically, the conjectured $c$-function can be written as
\bea
F = (-1)^{\frac{d-1}{2}} \log |Z| ~, & \qquad \qquad & d\ \mathrm{odd} ~,
\eea
where $|Z|$ is the $S^d$ partition function.
For interpolating solutions between AdS$_4$ and AdS$_2$, the free energy thus defined at the endpoint CFTs provides the right boundary value data to determine the constants of the bulk $c$-function~\cite{kleb,kleb1}.
In AdS$_2$, the finite part of the partition function has been shown to be nothing but the dimension of the Hilbert space of the dual CFT$_1$~\cite{2011cn}.
This corresponds to the Hilbert space of microstates of the single centered black hole.
In consequence, the boundary value data at the infrared fixed point is nothing but the entropy of the black hole or the entropy density of the black brane.

As monotonicity and extremization are consequences of the null energy condition, our proposal generalizes to all attractive flows in four dimensions including the non-supersymmetric, extremal ones.

\section{Determining the constants} \label{constants}

As a heuristic example, we present a model calculation to determine the constants of the $c$-function in an AdS$_5$ to AdS$_3$ interpolating black brane solution.
Consistent with the normalization of~\cite{sinha1}, the central charge of the four-dimensional CFT at the boundary is given in terms of the AdS$_5$ radius $L_\mathrm{UV}$ as
\be
c_\mathrm{UV}= \frac{\pi^2 L^3_\mathrm{UV}}{\ell_{5}^3} ~,
\ee
where $ \ell_{5}$ is the five-dimensional Planck scale.
As we are working in Einstein gravity, the $a$ and $c$ central charges are the same in the ultraviolet.
The central charge of the infrared CFT is given in terms of the AdS$_3$ radius $L_\mathrm{IR}$~\cite{bh}:
\be
c_\mathrm{IR}=\frac{3 L_\mathrm{IR}}{2 \ell_{3}}= \frac{3 b^2(r_h) L_\mathrm{IR}}{2 \ell_5} ~,
\ee
where $\ell_3$ is the three-dimensional Planck scale.
(In our conventions, $G_D = \ell_D^{D-2}$.)
Here, we can think of the planar part as a sphere of infinite radius.
The two constants in the $c$-function are then
\begin{equation}
\lambda = \frac{\H_\mathrm{UV}\, c_\mathrm{IR}- \H_\mathrm{IR}\, c_\mathrm{UV}}{\H_\mathrm{UV}- \H_\mathrm{IR}} ~, \qquad \qquad
\kappa = \frac{c_\mathrm{UV} - c_\mathrm{IR}}{\H_\mathrm{UV} - \H_\mathrm{IR}} ~.
\end{equation}
In the previous expression, $\H_\mathrm{UV}$ (respectively, $\H_\mathrm{IR}$) is the term in square brackets in~\eqref{c3} evaluated at $r=\infty$ ($r=r_h$).

Notice that, in so far as we can regard the central charge as encoding the number of degrees of freedom, at the infrared fixed point, which is a direct product of a two-dimensional space with AdS$_3$, the number of degrees of freedom, and hence, the central charge, is formally infinite, in cases where the two-dimensional space is non-compact.
In these cases, one simply quotients out the phase space by the volume of the two-dimensional space to get a finite volume density.
This leaves the phase space of AdS$_3$ diffeomorphisms which gives rise to a Virasoro $\times$ Virasoro algebra unchanged as the two spaces combine as a direct product to give the infrared geometry.
This is formally equivalent to taking the phase space to be the direct product of the space of diffeomorphisms of the AdS$_3$ factor and unit volume of the phase space of the trivial fiber, thus giving rise to a well-regulated finite $c$-function value at the infrared fixed point.

\section{Discussion} \label{disc}
The algorithm for determining the $c$-function is motivated by the general first order dynamics of the attractive flows which link the flow to a conservative vector valued function $f(r)$ over the scalars.
For a given diffeomorphism gauge, we can write down the geometrical quantity that encodes the function $f(r)$, and this allows us to deduce a $c$-function in that gauge.
The null energy condition, with a careful choice of null vector, implies monotonicity; extremization follows from the scalar flow in the bulk.

The physics at the ultraviolet and infrared fixed points of the boundary theory must be independent of the choice of scheme for the RG evolution.
Certainly, the $c$-function is not unique.
Based on~\cite{bm,hlp}, AdS/CFT equates diffeomorphism invariance in the bulk to scheme independence in the boundary theory~\cite{bgmr}.
We expect that other $c$-functions can be generated through general coordinate transformations in the spacetime.

Our results may be useful in investigating intermediate scaling vacua in attractor flows between critical points in moduli space.
The scaling regions between two fixed points of the flow are vacuum solutions of gauged supergravity, except that they break Lorentz symmetry and correspond to points in the RG flow on the boundary where the Hilbert space that contributes to the single centered black hole entropy contains Lorentz violating states.
A class of such solutions, namely the Lifshitz geometries, has been studied extensively as candidates for field theory states that exhibit quantum critical phase transitions~\cite{lif,lif1,bchno}.

In~\cite{dmpv}, bulk and boundary flows are shown to agree at the infrared fixed point via computation of three-point correlation functions of chiral primary operators in the two-dimensional CFT.
Since the $c$-function provides an affine parametrization of the flow globally, one use of our formalism is to generalize this result at first order away from the fixed point.
Our prescription allows the construction of $c$-functions in higher spin theories, which provide a proving ground for essential features of holography.
In these backgrounds, where the horizon for a black hole is a frame dependent concept~\cite{agkp}, the $c$-function may help define an unambiguous notion of an infrared fixed point that characterizes the spacetime.
As well, in the maps in~\eqref{cases}, we study black solutions either in CFT$_2$ or in CFT$_4$.
The existence of $c$-functions of the type we have discussed are characterized by embeddings of the Virasoro algebra in quantum field theories with $SO(d,2)$ conformal symmetry upon taking certain limits.
Also, it will be interesting to apply our procedure when there are higher derivative corrections~\cite{attr2,avo,higher,higher1,higher2,higher3}.
One more important direction of investigation is to determine how to reproduce our proposed $c$-function from holographic entanglement entropy calculation~\cite{sinha}.
We aim to explore these ideas in future work.

\acknowledgments
We thank Paolo Benincasa, Gabriel Cardoso, Robert de Mello Koch, Kevin Goldstein, Rob Myers, Shubho Roy, Joan Sim\'on, Aninda Sinha, and Jan Troost for deep discussions.
VJ is supported by the South African Research Chairs Initiative of the Department of Science and Technology and National Research Foundation.
He is grateful to KIAS for hospitality during the concluding stages of this work.
SSH, SN, and AVO are supported by SARChI, NRF, and the University Research Council.

\end{document}